\documentclass[twocolumn,usenatbib]{mnras}
\pdfoutput=1 %for arXiv submission
\usepackage{amsmath,amstext}
\usepackage[T1]{fontenc}
\usepackage{graphicx,graphics,amsmath}
\usepackage{amssymb}
\input{epsf}
\usepackage{graphicx}
\usepackage{color}

\def\be{\begin{equation}}
\def\ee{\end{equation}}
\def\beq{\begin{eqnarray}}
\def\eeq{\end{eqnarray}}

\title[Asymmetric accretion and thermal mountains]{Asymmetric accretion and thermal `mountains' in magnetized neutron star crusts.}
\author[N.~Singh et al.]{N.~Singh$^{1}$, B.~Haskell$^{2}$, D.~Mukherjee$^{3,4}$, T.~Bulik $^{1,2}$\\
$1$ Astronomical Observatory, University of Warsaw, Al. Ujazdowskie 4, 00-478 Warsaw, Poland\\
$2$ Nicolaus Copernicus Astronomical Center, Polish Academy of Sciences, ul. Bartycka 18, 00-716 Warsaw, Poland\\
$3$ Dipartimento di Fisica Generale, Universit\`{a} degli Studi di Torino, Via Pietro Giuria 1, 10125 Torino, Italy\\
$4$ INAF, Osservatorio Astrofisico di Torino, Strada Osservatorio 20, 10025 Pino Torinese, Italy}

\begin{document}
\maketitle

\begin{abstract}

Accreting neutron stars are one of the main targets for continuous gravitational wave searches, as asymmetric accretion may lead to quadrupolar deformations, or `mountains', on the crust of the star, which source gravitational wave emission at twice the rotation frequency. The gravitational wave torque may also impact on the spin evolution of the star, possibly dictating the currently observed spin periods of neutron stars in Low Mass X-ray Binaries and leading to the increased spindown rate observed during accretion in PSR J1023+0038.
Previous studies have shown that deformed reaction layers in the crust of the neutron star lead to thermal and compositional gradients that can lead to gravitational wave emission. However, there are no realistic constraints on the level of asymmetry that is expected.
In this paper we consider a natural source of asymmetry, namely the magnetic field, and calculate the density and pressure perturbations that are expected in the crust of accreting neutron stars.
In general we find that only the outermost reaction layers of the neutron star are strongly perturbed. The mass quadrupole that we estimate is generally small and cannot explain the increase of spin-down rate of PSR J1023+0038.
However, if strong shallow heating sources are present at low densities in the crust, as cooling observations suggest, these layers will be strongly perturbed and the resulting quadrupole could explain the observed spindown of PSR J1023+0038, and lead to observable gravitational wave signals from systems with higher accretion rates.

 \end{abstract}

\section{Introduction} 

The first detection of gravitational waves \citep{First} has opened the field of Gravitational Wave (GW) astronomy and the recent detection of a binary Neutron Star (NS) inspiral \citep{GW170817} has already begun to constrain the Equation of State (EoS) of dense matter \citep{GWEOS}. NSs are expected to be prolific GW emitters and we not only expect them to play a role in inspirals, but also to emit continuous gravitational waves (CWs), due to unstable modes or non-axisymmetric deformations (dubbed `mountains') that turn the star into a GW pulsar \citep{Riles}. 

This last scenario, in particular, has attracted much interest, as on the one side the neutron star crust is thought to be strong enough to support a significant deformation before cracking \citep{Haskell06, HorKadau, Baiko, JMD}, on the other accretion onto a magnetised star provides a natural mechanism to produce crustal asymmetries and source a deformation \citep{Bild98, UCB}. It has even been suggested that the spin rate of accreting NSs in Low Mass X-ray Binaries (LMXBs) may be set by the competition between the accretion torque that is spinning-up the star and the GW torque that removes angular momentum from the system \citep{PP78}. This would explain why we do not observe sub-millisecond pulsars, despite the Keplerian breakup frequency of a NS $\nu_k$, being above $\nu_k\approx 1200$ Hz for any physically realistic model that is causal in the core and includes a crust \citep{Haskell18}.

A recent analysis of the spin-distribution of accreting NSs in LMXBs by \citet{Patruno17} has also revealed that the distribution is bimodal, with a broad population appearing at lower frequencies (and consistent with the observed population of millisecond radio pulsars, of which the LMXBs are thought to be the progenitors \citet{Alpar82}) and a much more narrowly distributed population of rapidly rotating neutron stars with $550$ Hz $\lesssim \nu\lesssim 720$ Hz. This feature, in particular, suggests a torque that scales strongly with frequency and is difficult to explain simply involving accretion torques, but can be naturally explained if GW torques are acting on the system \citep{Gittins}.

One of the pulsars in the fast population, PSR J1023+0038, has attracted particular attention. This is a transitional pulsar, that transitions from being accretion powered in outburst to radio emission during quiescence \citep{J10}. It has been possible to time this pulsar both in the X-ray and radio phases, and measurements of the spin-down rate have revealed that the NS is spinning down $\approx27\%$ faster when it's accreting \citep{Jao16}. This result is somewhat unexpected, as standard accretion models can predict spin-down during accretion, but generally require some fine-tuning for the spin-down rate in both phases to differ so little \citep{Jao16}. Gravitational wave emission, however, offers a scenario that can explain the slight increase in spin-down rate quite naturally if a mountain is being built on the crust during the accretion outburst \citep{Haskell17}.

If light elements are accreted at the surface, they are subsequently pushed deeper into the crust when new material is added, and here they undergo several reactions, including pycno-nuclear reactions and electron capture, that change the composition and release heat locally \citep{Haensel90}. This deep-crustal heating is responsible for heating X-ray transients, which then cool during quiescence as has been observed in many systems \citep{Natalie}. If part of the reaction layers are not exactly axisymmetric,  temperature and composition gradients can source a `mountain' and lead to GW emission \citep{Bild98,UCB}. 

The main parameters that determines the size of the mountain are the total accreted mass, which depend on the accretion rate and outburst duration, and the amount of asymmetry in the heat release. For example \citet{Haskell17} found that if $\approx 2\%$ of the heat released is quadrupolar, then a large enough mountain, that explains the observed spindown rate, can be built on PSR J1023+0038.

How much of the heat emission will be quadrupolar, or how quadrupolar the surfaces of equal composition will be, is however, generally unknown. One can put an upper limit of around $50 \%$ on the ratio between quadrupolar and spherical components of the heat deposition from the fact that no pulsations are visible in quiescence with an amplitude larger than $\approx50\%$, while if there were quadrupolar heat deposition, the heat would diffuse out in quiescence and cause pulsations at twice the rotation frequency with similar amplitude (see \citet{Haskell15} for a detailed analysis). 
There is, ultimately, no realistic estimate of how large the quardupolar component is expected to be, thus limiting the predictive power of this model. 

In this paper we consider a natural source of asymmetry, namely the magnetic field. We model accretion onto the polar cap of a neutron star numerically with the code of \citet{code12}, and calculate how distorted the density and pressure profiles will be in the outer layers of the star. We then extrapolate to higher densities, to calculate the quadrupolar component of the temperature perturbation due to reactions in the outer crust up to neutron drip density ($\rho\approx 10^{11}$ g\;cm$^{-3}$). From this we use the results of \citet{UCB} to estimate the induced ellipticity and gravitational wave signal. In general we find that only relatively small ellipticities can be produced by this mechanism, smaller than could be detected by Advanced LIGO and Virgo and smaller than what would be required to explain the additional spindown  of PSR J1023+0038.

\section{Deep crustal heating}

A NS in an LMXB accretes matter, generally hydrogen or helium, from the companion, and as accretion proceeds these light nuclei are pushed down to higher densities, and can undergo a number of reactions, including electron captures and pycno-nuclear fusions, which produce heavier nuclei and push the composition towards equilibrium \citep{Haensel90}.
This process releases a few MeV per accreted baryon of heat in the crust, which is thought to power the quiescent luminosity of NS in LMXBs and can explain the observed cooling of transient LMXBs \citep{Natalie}.
If the reaction layers are not perfectly axisymmetric, gradients in temperature and composition resulting from asymmetric accretion can source a quadrupolar deformation, sustained by the rigidity of the crust \citep{Bild98}. \citet{UCB} examined this problem in detail and calculated the quadrupole induced by reactions in the outer crust. Their results, assuming a quadrupolar  temperature increase $\delta T_{q} $, can be approximated as:
\be
Q_{22}=1.5\times10^{35}R^{4}_{10}\left(\frac{\delta T_{q}}{10^{5}}\right) \left(\frac{E_{th}}{30 \;\rm{MeV}}\right)^{3} \rm {g\;cm^{2}} \label{quadro1}
\ee
where $R_{10}$ is the radius in units of 10 km, $E_{th}=\mu_e$, with $\mu_e$ the electron chemical potential, is the threshold energy for the pycnonuclear reactions responsible for deep crustal heating in a given layer. Since the required quadrupole $Q_{22}$ to explain the behaviour of PSR J1023+0038 is $4.4\times10^{35}I_{45}$ g $\rm {cm^{2}}$ \citet{Haskell18}, where $I_{45}$ is the moment of inertia of the NS in units of $10^{45}$ g $\rm {cm^{2}}$, from eq. (\ref{quadro1}) we see that the required quadrupolar temperature variation for this star is
\be
\delta T_{q}\approx 2.9\times10^{5}R^{-4}_{10}I_{45}\left( \frac{E_{th}}{30\;\rm{MeV}}\right)^{-3}\;\;\mbox{K}\; .\label{quadro2}
\ee
Assuming that deep crustal heating is the only source of temperature increase, the total increase in temperature is given as \citet{Rutledge}:
\be
\Delta T \approx 10^{2}C_{k}^{-1}p^{-1}_{30}Q_M\Delta M_{21}\;\;\mbox{K}\label{aumento}
\ee
where $C_{k}$ is the heat capacity in units of the Boltzman constant per baryon, $p_{30}$ is the pressure in units of $10^{30}$ erg/$\rm{cm^{3}}$, $Q_{M}$ is the heat released locally by the reactions per accreted baryon, in units of MeV, and $M_{21}$ is the accreted mass in units of $10^{21}$ g.

To obtain an estimate for PSR J1023+0038 \citet{Haskell17} considered a simplified model with only the contribution due to the most energetic reactions close to the neutron drip layer, at $\rho\approx 4\times 10^{11}$ g\;cm$^{-3}$, although note that for accreted crusts this may be shifted to higher densities depending on the composition of the ashes of X-ray bursts, as discussed by \citealt{Chamel15}.
We have  used $\Delta M_{21}=0.1$ following  \citet{Haskell17}, which is reasonable for a month of accretion.
In this case, taking $E_{th}=30$ MeV, $Q_M=0.5$ MeV, $C_k=10^{-6}$ and $p_{30}=1$, one obtains
\be
\frac{\delta T_q}{\Delta T}\gtrsim 0.03\;\;\mbox{(at neutron drip)}\label{quadfrac}
\ee

Naturally a more detailed analysis should consider the contribution due to all reaction layers, which will add up to contribute to the total quadrupole, thus reducing the estimate in (\ref{quadfrac}). Taking the results of \citet{Haskell17} and summing the contribution of all layers in the outer crust, as suggested by \citet{UCB}, for a linearized model, the total quadrupole generated by $i$ reactions in different layers will be given as:
\be
Q_T\approx  1.5\times 10^{32} \Delta M_{21} R_{10}^4\sum_ i \frac{(Q_M)_i  (E^3_{30})_i}{(p_{30})_i (C_k)_i} \left(\frac{\delta T_q}{\Delta T}\right)_i \label{total}
\ee
where $E_{30}$ is $E_{th}$ in units of $30$ MeV, and all quantities with the subscript $i$ must be calculated at the pressure and density corresponding to the reaction layer $i$.

It remains an open issue whether deformations of order $\delta T_q/\Delta T\approx 0.01$ can occur in the crust of an accreting NS, and without an estimate of this quantity it is impossible to obtain a meaningful theoretical estimate of the quadrupole from (\ref{total}), let alone compare this to observations.

There is, however, a natural source of asymmetry in the system, as the NS is magnetised, and unless the field is too weak, matter is accreted onto the polar caps and then spreads due to lateral pressure gradients, that are opposed by magnetic stresses. This can lead to the creation of magnetic mounds in accreting systems \citep{Payne1, Payne2, DipanjanRev}, and significantly deform the field structure, leading to burial of the dipolar component \citep{Shib89}. In fact, for high magnetic fields and accretion rates, the magnetically confined mountain may itself be large enough to lead to a detectable gravitational signal \citep{Max, Haskell15}. We will see, however, that for the weaker fields we consider the magnetic contribution to the ellipticity is generally much smaller than the `thermal' contribution due to deep crustal heating.

In summary our strategy consists of splitting the full problem, which would include contributions due to magnetic stresses, elasticity and the readjustment of the reaction layers due to the non-barotropic nature of a realistic EoS in the crust, into following steps. First we calculate  barotropic magnetic equilibrium to estimate the temperature asymmetries in the crust of the accreting star. Then we use these to calculate the source terms that lead to the readjustment of the capture layers and then using the results of \citet{UCB}, we obtain the elastic response of the crust.

\section{Temperature asymmetries}
To calculate the temperature asymmetries in the crust we consider a sequence of static equilibria of a mound of accreted matter confined by a magnetic field in the neutron star outer layers. 
The equilibrium solution is constructed by numerically solving the Grad-Shafranov (GS) equation using the numerical GS solver developed by \citet{code12}, described in detail in Section \ref{GS}. As in the original algorithm of the GS solver, the effects of continued accretion are not taken into account. Although it has been demonstrated that pressure driven MHD instabilities \citet{inst1,inst2} can occur beyond a threshold mass of the magnetically confined mounds, \citet{Vigelius} have shown that the instabilities may saturate to a new equilibrium configuration. For analytical simplicity, we perform the analysis in this paper on the equilibrium solutions obtained from the GS solver, without exploring the stability of the mound, to obtain an upper limit on the deformations and the resultant gravitational wave signals. We then, use the equilibrium solution to estimate the asymmetries in the crust which is described in Section \ref{assym}

\subsection{Formulation of Grad-Shafranov equation} 
\label{GS}
To simulate the mass and magnetic field configuration on the surface of a neutron star in an LMXB we consider the equilibrium profile of a `mountain' on the surface of a star accreting matter from a disc, where the magnetosphere cuts off the accretion disk at the Alfven radius, the distance from the star where the energy density of the stellar magnetic field is balanced with the energy density of the inflowing matter and the matter is accreted onto the poles. As we shall see in the following this allows us to define the extent of the polar cap on which matter is deposited by accretion. More detailed treatments of the inner disc radius are possible \citep{rad1, rad2, rad3, rad4, rad5}, but generally the position of the inner edge of the disc $r_{inner}$ is found to be proportional to the Alfven radius $r_A$, such that $r_{inner}=\xi r_A$, with $0.4\lesssim \xi\lesssim 1$ \citep{Dangelo17}. For simplicity, and given the other sources of uncertainty in our treatment, we will thus simply take $r_{inner}=r_A$.

The initial magnetic field is dipolar and the polar caps are assumed to be axisymmetric about the magnetic axis $\textbf{z}$.  For zero toroidal magnetic field in an axisymmetric system $\left(r,\theta,\phi \right)$, in ideal magnetohydrodynamics (MHD) the poloidal magnetic field is given as:
\be
 \textbf{B}_{p}=\frac{\boldsymbol{\nabla}\psi(r,\theta)\times \boldsymbol{\hat{\phi}}}{r\sin\theta}\label{poloidalmagfld}
\ee
where $\psi$, the flux function is given as:
\be
\psi=r\sin\theta A_{\theta}\label{fluxfunc}
\ee
where, $A_{\theta}$ is the azimuthal component of the vector potential and the system is symmetric with respect to azimuthal coordinate $\phi$. The azimuthal component of the current is then,
\be
\textbf{j}_{\phi}=-\frac{c}{4\pi}\frac{\Delta^{2}\psi}{r\sin\theta}
\ee
\be
\Delta^{2}=\frac{\partial^{2}}{\partial r^{2}}+\frac{\sin\theta}{r^{2}}\frac{\partial}{\partial\theta}\left(\frac{1}{\sin\theta}\frac{\partial}{\partial\theta}\right) 
\ee
where, $\Delta^{2}$ is the Grad-Shafranov operator in spherical coordinates.
For a system in static equilibrium, the Euler equation is,
\be
\boldsymbol{\nabla} p +\rho\boldsymbol{\nabla}\Phi=\frac{\textbf{j}\times \textbf{B}}{c}=\rho\boldsymbol{\nabla} F\label{euler}
\ee
where $\Phi$ is the gravitational potential and $\boldsymbol{\nabla} F = \frac{\boldsymbol{\nabla} p}{\rho} + \boldsymbol{\nabla}\Phi$, so F is a function constant along a flux surface. For a spherical coordinate system and Newtonian gravity with constant acceleration,  
\be
\boldsymbol{g}=-1.86\times10^{14}\left(\frac{M_{s}}{1.4M_{0}}\right)\left( \frac{R_{s}}{10\rm km}\right)^{-2} \rm cm s^{-2} \hat{\textbf{r}}
\ee
assuming $p=f(\rho)$, the density profile is obtained by integrating eq.(\ref{euler}) to get,
\be
\frac{1}{\rho}\frac{df}{d\rho}\boldsymbol{\nabla}\rho=\boldsymbol{\nabla}(F(\psi)-\Phi)\label{densprof}
\ee
For a polytropic equation of state $p=K\rho^{\Gamma}$ eq.(\ref{densprof}) gives,
\be
\rho=\left( \frac{g(\Gamma-1)}{\Gamma K}\right)^{\frac{1}{\Gamma-1}}\left( r_{0}(\psi)-r\right) ^{\frac{1}{\Gamma-1}}\label{mounddensity}
\ee
where $r_{0}=F(\psi)/g$ is an arbitrary mound height function, the choice of which gives the shape of the mound. The constant of integration is taken to be zero. Substituting eq.(\ref{mounddensity}) in eq.(\ref{euler}) gives,
\be
\Delta^{2}\psi=-4\pi r^{2}\sin^{2}\theta\rho\frac{dF}{d\psi}=-4\pi r^{2}\sin^{2}\theta\rho g\frac{dr_{0}}{d\psi}\label{GradShafranov}
\ee
Eq.(\ref{GradShafranov}) is the Grad-Shafranov equation in spherical coordinates. The above equation has been solved using a parabolic mound height function of the form:
\be
r_{0}(\psi)=R_{s}+r_{c}\left( 1-\left(\frac{\psi}{\psi_{A}}\right)^{2} \right)\;\;\mbox{parabolic profile}\label{parbprof}
\ee ,
and hollow mound height function given as:
\be
r_{0}(\psi)=R_{s}+\frac{r_{c}}{0.25}\left( 0.25 -\left(\frac{\psi}{\psi_{A}}-0.5\right)^{2} \right)\;\;\mbox{hollow profile}\label{hollowprof}
\ee
which accounts for the fact that mass loading will occur over a finite range of accretion disc radii, as discussed by \citet{inst1}.
Here $\psi_{A}$ is the flux function at the Alfven radius and $r_{c} $ the cutoff chosen for the mound height. The density is determined by mound height profile and eq.(\ref{mounddensity}).

Specifically we take polytropic equation of state (EOS) of the form : $p(\rho)=K\rho^{\Gamma}$ where $\Gamma $ is the adiabatic index and $K$ is measured in cgs units (dyn $\rm g^{-\Gamma}\rm cm ^{3\Gamma - 2}$). The values of $K = 5.4 \times10^{9}$ and $\Gamma = 5/3 $ are chosen to crudely approximate to the density regimes of interest in the crust i.e degenerate neutron gas $10^{12} \leq \rho/$(g $\rm cm^{-3} )\leq 10^{16} $. The neutron star parameters are taken as $M_{s} = 1.4 M_{\odot} $, $R_{s} = 10 \rm km $, where $M_{s}$ and $R_{s}$ are the mass and radius of the neutron star and $M_{\odot}$ is the solar mass.  We note that our choice of EOS is mainly dictated by the desire to simulate high densities and models with large values of the accreted mass. Our choice ensures this (see \citet{Max} for a more detailed discussion) but is not entirely consistent as, in practice, models used for our extrapolation do not reach such high densities. A consistent model should consider the pressure as mainly due to degenerate electrons rather than neutrons. We computed equilibria using various EOS and found that a degenerate electron EOS is suitable for modelling the accreted matter only in the upper layers and for small accreted mass. To explore the maximum limit of the accreted mass sustainable inside the crust we use the non-relativistic degenerate neutron EOS to model the possible confinement of denser matter, keeping in mind that it is likely to provide an upper limit to the true value of the quadrupole that we calculate, and that a more realistic EOS should be considered in future work.

The Alfven radius $r_A$ is deduced by equating ram pressure to magnetic pressure and can be expressed as \citet{1977ApJ...215..897E} 

\begin{multline}
r_{A}=  3.53\times 10^{3}\left(\frac{B_{s}}{10^{12}\;\rm G}\right)^{4/7}\left(\frac{R_{s}}{10\;\rm km}\right)^{12/7} \\ \times \left(\frac{\dot{M}}{10^{-9}\;M_{\odot}\rm yr^{-1}}\right)^{-2/7}\left(\frac{M_{s}}{1.4\;M_{\odot}}\right)^{-1/7} \; \rm km\label{Alfvenradius}
\end{multline}
The maximum size of the polar cap is determined by the field lines from the Alfven radius on to the neutron star surface. The polar cap radius $ R_{p}$ can be expressed in terms of Alfven radius $ r_{A}$ as :
\be
R_{p}= \left(\frac{R_{s}}{r_{A}}\right)^{1/2}R_{s}
\ee
If $R_{p},\; \theta_{p} $ are the polar cap radius and the opening polar cap angle respectively, then $\theta_{p}$ is given as:
%$ R_{p}$ is given as:

%\be
%R_{p}= R_{s}\sin\theta_{p}\label{polarradius}
%\ee

\be
\theta_{p} = \sin^{-1}\left(\frac{R_{p}}{R_{s}}\right)
\ee

\subsection{Estimate of asymmetries in the crust}
\label{assym}

\begin{table*}
\caption{
Reactions in the crust of an accreting NS assuming
that the ashes of X-ray bursts consist of pure $^{56}{\rm Fe}$ using the EDF BSk21 
as described in \citet{Fantina18}. We provide pressure $P$, density $\rho$, threshold energy $E_{th}$ and heat release per accreted baryon $Q_M$. For completeness we also provide the free neutron fraction $X_n$ and the density jump in the layer $\Delta \rho/\rho$.}
\label{deep}
\begin{center}
\begin{tabular}{lllllrr}
\hline \hline \noalign{\smallskip} $P$ & $\rho$ & Reactions & $X_n$ &
$\Delta \rho/\rho$  &$E_{th}$ &  $Q_M$
\\ (dyn~cm$^{-2}$) & (g~cm$^{-3}$) & & &  & (MeV) 
&(keV)  \\ \hline \noalign{\smallskip}
$6.48\times 10^{26}$ &
$1.38\times 10^9$ & $^{56}\mathrm{Fe}\rightarrow
{}^{56}\mathrm{Cr}-2e^-+2\nu_e $ & 0 & 0.08 & 4.47 & 37.0\\
%& & & & \\
$1.83\times 10^{28}$ & $1.81\times 10^{10}$
&$^{56}\mathrm{Cr}\rightarrow {}^{56}\mathrm{Ti}-2e^-+2\nu_e$ & 0 & 0.09 & 10.22& 41.2\\
% & & & & \\
$1.06\times 10^{29}$ & $7.37\times 10^{10}$
 & $^{56}\mathrm{Ti}\rightarrow {}^{56}\mathrm{Ca}-2e^-+2\nu_e$ & 0 & 0.10 & 15.83 & 39.1\\
% & & & &  \\
$3.43\times 10^{29}$ & $1.96\times 10^{11} $
 & $^{56}\mathrm{Ca}\rightarrow
 ^{56}\mathrm{Ar}-2e^-+2\nu_e$ & 0 & 0.11 & 21.22 & 8.1\\
% & & & & \\
$8.75\times 10^{29}$ & $4.38\times 10^{11} $
 & $^{56}\mathrm{Ar}\rightarrow {}^{55}\mathrm{Cl}+n-e^-+\nu_e$ & 0 & 0.06 & 26.55 & 0\\
$9.40\times 10^{29}$  &   $4.79\times 10^{11}$   &
$^{55}\mathrm{Cl}\rightarrow {}^{53}\mathrm{S}
+\Delta N\cdot n-e^-+2\nu_e$   &   0.05   &   0.06&27.04 &0\\
%%&&&&&\\
$1.18\times 10^{30}$  & $6.04\times 10^{11}$   &   ${}^{53}\mathrm{S}\rightarrow
{}^{48}\mathrm{Si}+\Delta N\cdot n-2e^-+2\nu_e$  &  0.15  &0.14  & 28.57 &45.0\\
%%%&&&&&\\
\hline
\noalign{\smallskip}
$2.54\times 10^{30}$   &   $1.22\times 10^{12}$   &
${}^{48}\mathrm{Si}\rightarrow {}^{30}\mathrm{O}+\Delta N\cdot n-6e^-+2\nu_e
$   &  &  && \\
%%&&&&&\\
\noalign{\smallskip}
&&
$ {}^{30}\mathrm{O}+{}^{30}\mathrm{O}\rightarrow {}^{51}\mathrm{Si}+\Delta N\cdot n-2e^-+2\nu_e$&0.54& 0.68 &32.64&908.1\\
\hline
%%%&&&&&\\
\noalign{\smallskip}
%\hline
$5.78\times 10^{30}$   &   $3.73\times 10^{12}$   &
${}^{53}\mathrm{Si}\rightarrow {}^{32}\mathrm{O}+\Delta N\cdot n-6e^-+2\nu_e
$   &  &  && \\
%%&&&&&\\
\noalign{\smallskip}
&&
${}^{32}\mathrm{O}+{}^{32}\mathrm{O}\rightarrow {}^{62}\mathrm{S}+\Delta N\cdot n$&0.72& 0.23 &35.47&355.9\\
\hline
\noalign{\smallskip}
 $8.69\times 10^{30}$   &   $6.16\times 10^{12}$ &${}^{64}\mathrm{S}\rightarrow {}^{57}\mathrm{Si}+\Delta N\cdot n-2e^-+2\nu_e$ &  
 0.74  &   0.03&37.74&3.5\\
%%%&&&&&\\
\hline
\noalign{\smallskip}
$3.20\times 10^{31}$   &   $1.65\times 10^{13}$   &
${}^{65}\mathrm{Si}\rightarrow {}^{40}\mathrm{O}+\Delta N\cdot n-6e^-+2\nu_e
$   &  &  && \\
%%&&&&&\\
\noalign{\smallskip}
&&
$ {}^{40}\mathrm{O}+{}^{40}\mathrm{O}\rightarrow {}^{76}\mathrm{S}+\Delta N\cdot n$&0.83& 0.05 &43.8&98.2\\
\hline
\noalign{\smallskip}
$1.85\times 10^{32}$  & $7.26\times 10^{13}$   &   ${}^{91}\mathrm{S}\rightarrow
{}^{86}\mathrm{P}+\Delta N\cdot n-e^-+\nu_e$  &  0.81  &0.006  & 69.10 &0\\

%&&&&&\\
\hline
\hline
\end{tabular}
\end{center}
\end{table*}
%%%%%%%%%%%%%%%%%%%%%%

In the spherical coordinate system $\left(r, \theta, \phi \right)$ used here, with $r$ the distance from the center of the star, $\theta$ the angle from the magnetic axis, and $\phi$ the azimuthal angle, the system is symmetric with respect to the azimuthal coordinate $\phi$. The density values can be written as $\rho\left(r, cos\theta\right)$, and expanded in terms of standard spherical harmonics with $m=0$ ($Y_{l0}$) in the form:
\begin{equation}
\rho\left(r, cos\theta\right)=\sum_l \rho_{l}(r)Y_{l0} \label{expand}
\end{equation}
and each component is calculated as :
\begin{equation}
\rho_l(r)=2\pi \int \rho(r,cos\theta) Y_{l0}d\cos\theta
\end{equation}
As we are interested in the quadrupolar deformation which gives the leading order contribution to the gravitational wave flux, we focus on the $l=2$ coefficients and calculate the values for the ratio, $\rho_{2}/\rho_{0}$ for each layer in the mound. This quadrupolar variation in density generated due to magnetic stress is then used to evaluate quadrupolar variation in the temperature for the estimation of the quantity $\delta T_{q}/\Delta T$, from equations (\ref{estimo1}), (\ref{estimo2}) and (\ref{estimo3}).

We note that as the numerical setup of the system is symmetric with respect to the azimuthal coordinate $\phi$, a direct expansion in Legendre polynomials may have been more natural. However we choose to expand in spherical harmonics as the true quantity that is needed to estimate the quadrupolar deformation in eq. (\ref{quadro2}) is the $l=2, m=2$ component of the temperature perturbation. We make the assumption as in \citet{Hask08} that $\delta T_{22} \approx \delta T_{20},\, \delta \rho_{22} \approx \delta\rho_{20}$, the subscripts being the $l$ and $m$ values respectively. We have chosen to do so since for slow rotation of the star the two quantities differ only by a geometric factor due to the misalignment of the rotation axis (which sets the spherical coordinate basis with respect to which the spherical harmonics are defined) and the magnetic axis. The estimate made here thus gives the $l=2,m=0$ component of the temperature perturbation rather than the $l=2, m=2$ component, but allows us to solve for more tractable model in which the magnetic and rotational axis are aligned \citep{BG96,Hask08}.

Keeping in mind the approximation $\delta T_{22}\approx\delta T_{20}$, to estimate the quadrupolar perturbation in the temperature, let us consider linear perturbations of the pressure, around a spherically symmetric background value $p_0(r)$ i.e,
\begin{equation}
p(r,\theta)= p_{0}(r) + \delta p(r,\theta) + O\left( \delta p\right)^{2}
\end{equation}
where $\delta p (r,\theta)$ can also be written in terms of density perturbations as,
\begin{equation}
\delta p (r,\theta) = \frac{\partial p_0}{\partial \rho_{0}}\delta \rho (r,\theta)
\end{equation}
If we assume an expansion in spherical harmonics for the perturbations also, such that , e.g.
\be
\delta p(r,\theta)=\sum_l \delta p_l(r) Y_{l0}(\theta)
\ee 
we can use this linear perturbation in eq. (\ref{aumento}) to obtain the value for quadrupolar temperature variation as,
\begin{equation}
\delta T_{q} = -10^{2}C_{k}^{-1}p^{-1}_{30} Q_M\Delta M_{21} \left[\left(\frac{\delta p_{30,q}}{p_{30}} \right)+\left(\frac{\delta C_{k,q}}{C_{k}} \right)\right]
\end{equation}
or, equivalently:
\be
\frac{\delta T_q}{\Delta T}=- \left[\left(\frac{\delta p_{30,q}}{p_{30}} \right)+\left(\frac{\delta C_{k,q}}{C_{k}} \right)\right]\label{estimo1}
\ee
where with the subscript $q$ we are now indicating the $l=2$ coefficient of the expansion.
We assume $Q_M$ to be constant which is a good approximation on the timescale of an accretion outburst as $Q_M$ can only change on much longer evolutionary timescales when the star first starts accreting and the cold catalyzed crust is replaced by accreted material.
Assuming a polytropic equation of state we can write 
\be
\left(\frac{\delta p_{30,q}}{p_{30}} \right)=\Gamma\frac{\delta\rho_q}{\rho}\label{estimo2}
\ee
At the densities of interest the main contribution to the heat capacity is that of the ions in the lattice \citep{POTReview15}. We therefore take \citep{Chong94}:
\be
C_k=\frac{3}{A} f \left(\frac{T}{\Theta_D}\right)\label{Ck}
\ee
where $\Theta_D=3.48\times 10^3\rho^{1/2}Z/A$ K is the Debye temperature at a given density $\rho$, is function of the proton number $Z$ and the atomic number $A$. The Debye function $f(x)$, with $x=T/\Theta_D$, can be approximated as \citep{VR91}

\begin{align}
    f(x) = \left\{ \begin{matrix}
          0.8\pi^4 x^3 & x\leq 0.15 \\
         1-0.05 x^{-2} & x\geq 0.4 \\
          1.7 x + 0.0083 & \mbox{ otherwise}
         \end{matrix} \right.
  \end{align}

Perturbing eq. (\ref{Ck}) we have,
\be
\delta C_{k,q}=\frac{\partial C_{k,q}}{\partial\rho}\delta\rho_q+\frac{\partial C_{k,q}}{\partial T}\delta T_q=\frac{3}{A}\frac{d f}{d x}\left(\frac{\delta T_q}{\Theta_D}-\frac{x}{2}\frac{\delta\rho_q}{\rho}\right)\label{estimo3}
\ee
so that for the temperatures of interest i.e.  $T \lesssim 10^7$K or $x\leq 0.4$ we have
\begin{align}
\frac{\delta C_{k,q}}{C_k}\approx \left\{ \begin{matrix}
	-\frac{1}{2}\left(\frac{\delta\rho_q}{\rho}\right)+\left(\frac{\delta T_q}{ T}\right) &\;0.15 \ensuremath < x\leq 0.4 \\
	-\frac{3}{2}\left(\frac{\delta\rho_q}{\rho}\right)+3\left(\frac{\delta T_q}{ T}\right) &\;x\leq 0.15 
\end{matrix} \right.\label{deltaT}
\end{align}

In the following we will assume that the crustal heating is strong enough in the capture layers, so that $T\approx \Delta T(l=0)$ \citep{Rutledge}. 
We obtain $\frac{\delta T_q}{\Delta T}$ by combining eq.  \ref{deltaT}, \ref{estimo1} and \ref{estimo2}. Thus the spherically symmetric heat increase obtained from equation (\ref{aumento}) sets the background temperature of our model, and also identify the background $\rho$ with $\rho(l=0)$ extracted from our simulation.
Thus the main assumption that allows us to obtain the result is that ratio $\delta T_q/\Delta T$ can be expressed in terms of the ratio of $l=2$ and $l=0$ component of mass density perturbation $\rho_2/\rho_0$, which is generated due to the confinement of accreted material by magnetic stress produced by the bending of field lines. It is based on the fact that the temperature of the crust in many cases is set by crustal heating \citep{Rutledge}. The only additional ingredient we need for our calculation of the quadrupole is the density $\rho$, pressure $p$, threshold energy $E_{th}$ and energy release $Q_M$ of the deep crustal heating reactions in the outer crust. We take the recent results of \citet{Fantina18}, and the relevant parameters can be found in table \ref{deep}.

\section{Results}

To determine the quadrupolar density deformations we computed equilibria for varying values of accreted mass $\Delta M$ and magnetic field strength $B$, for two different values of the accretion rate, a low value of $\dot{M}=6\times 10^{-13} M_\odot\;\rm yr^{-1}$ which would be appropriate for a system such as J1023+0038 in which continuous gravitational wave emission was suggested \citep{Haskell17}, and a high value of $\dot{M}=3\times 10^{-8}M_\odot\;\rm yr^{-1}$, more appropriate for persistent sources accreting close to the Eddington limit, which are likely to be the best targets for gravitational wave searches \citep{Haskell15}.

We compute the density profile for each model and expand it as equation \ref{expand}, after which we compute the ratio $ \rho_2/ \rho_0$. Assuming that the effect of the magnetic field is weak compared to gravity, so as to maintain a spherically symmetric background density, we make the approximation that $\delta\rho_q/\rho_0\approx \rho_2/ \rho_0$, which allows us to determine $\delta T_q/\Delta T$ from equation (\ref{estimo1}). 

\begin{figure*}
\includegraphics[width=\columnwidth]{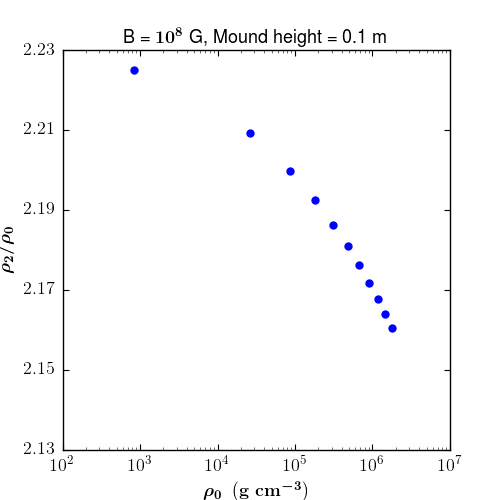}\includegraphics[width=\columnwidth]{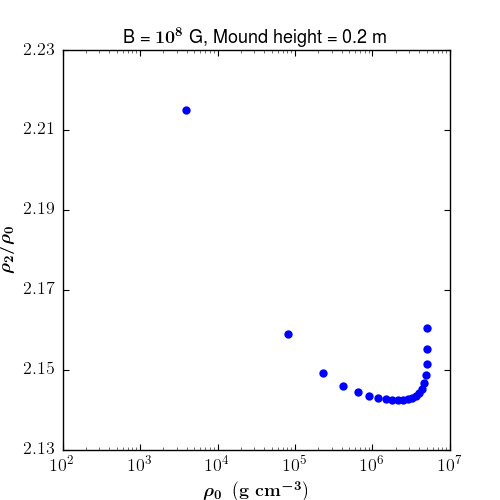}
\caption{Plots of the ratio $\rho_2/\rho_0$ versus the spherical density distribution $\rho_0$ for two parabolic (full) mound models with $B=10^8$ G and other parameters as in table \ref{tabexp1}. The ratio is plotted for two heights of the mound, 0.1 m on the left and 0.2 m, which is the last stable model, on the right. As can be seen for lower mound heights the $\rho_2/\rho_0$ decreases roughly linearly with density, but for the highest mound models (and thus largest accreted masses) the relation is highly non linear, and we thus do not attempt to extrapolate these models to higher densities in the crust.}\label{ratioP}
\end{figure*}

\begin{figure*}
\includegraphics[width=1\columnwidth]{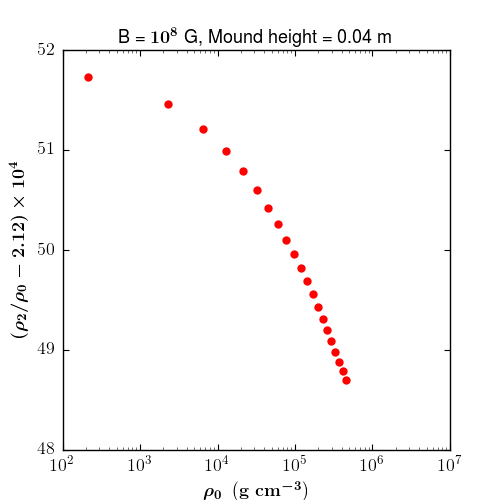}\includegraphics[width=1\columnwidth]{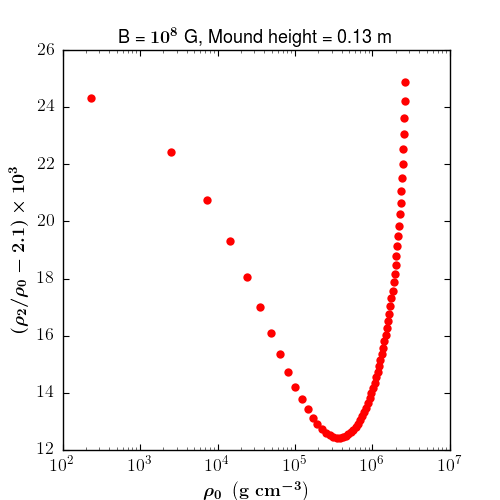}
\caption{Plots of the ratio $\rho_2/\rho_0$ versus the spherical density distribution $\rho_0$ for two hollow mound models with $B=10^8$ G and other parameters as in table \ref{tabexp2}. The ratio is plotted for two heights of the mound, 0.04m on the left and 0.13m, which is the last stable model, on the right, as in figure \ref{ratioP}. Also in this case the relation between $\rho_2/\rho_0$ and $\rho_0$ appears linear for lower accreted masses.}\label{ratioH}
\end{figure*}

In tables \ref{tabexp1}, \ref{tabexp2} and \ref{tabexp3} we show example of the results for an initial magnetic field strength of $B=10^8$ G (appropriate for J1023+0038) for both the parabolic (filled) and hollow profile. Examples of the ratio $\rho_2/ \rho_0$ are plotted in figures \ref{ratioP} and \ref{ratioH}, for different values of the accreted mass (or alternatively mound height), and extensive tests have been run for different values of the magnetic field strength. Table \ref{tabhighMB} shows the results of the simulation for a magnetic field strength of $B=10^{10}$ G and accretion rate $\dot{M}= 3\times 10^{-8} M_\odot\;\rm yr^{-1}$, (approximately the Eddington limit), for a parabolic mound profile.


\begin{table*}
\caption{Results of our simulations for a parabolic mound profile with a magnetic field $B= 10^{8}\rm G$, Accretion rate: $6\times10^{-13} M_\odot\;\rm yr^{-1}$. The polar spread corresponding to this field strength is 14.9 degrees and the maximum mound height that can be obtained is of 0.2 m, after which the models are unstable. We also compute the ellipticity for this model, although the reader should keep in mind that this is purely the contribution induced by the magnetically supported mound, and does not include contributions due to crustal reactions. The maximum base density is the maximum that is obtained scanning over all grid points of a simulations, while the maximum and minimum of $\rho_0$ represent the spherical $l=0$ component of the multipole expansion of the density.}
\label{tabexp1}%\label{density10^8}
\begin{center}
\begin{tabular}{c|c|c|c|c|c|c|c}
\hline \hline \noalign{\smallskip}
Mound Height & Max base density & Accreted mass & $\rho_2/\rho_0$ (Max)& $\rho_2/\rho_0$ (Min)& $\rho_0$ (Max)& $\rho_0$ (Min)& Ellipticity\\
 (m) & ($\rm g\;cm^{-3}$) & ($M_\odot$) &  &  & ($\rm g\;cm^{-3}$) & ($\rm g\;cm^{-3}$)&\\
\hline

0.05 & $1.808\times 10^{6}$ & $2.44\times 10^{-15}$ & 2.2251 & 2.1605 & $6.323\times 10^{5}$ & $2.933\times 10^{2}$ & $4.93\times 10^{-17}$\\
0.10 & $5.114\times 10^{7}$ & $1.23\times 10^{-14}$ & 2.2249 & 2.1605 & $1.788\times 10^{6}$ & $8.415\times 10^{2}$ & $2.44\times 10^{-16}$\\
0.15 & $9.395\times 10^{7}$ & $3.48\times 10^{-14}$ & 2.2237 & 2.1605 & $3.286\times 10^{6}$ & $1.685\times 10^{3}$ & $7.24\times 10^{-16}$\\
0.20 & $1.446\times 10^{8}$ & $9.34\times 10^{-14}$ & 2.2150 & 2.1425 & $5.104\times 10^{6}$ & $3.930\times 10^{3}$ & $2.46\times 10^{-15}$\\
0.25 & unstable & & & & & &\\
\hline \hline \noalign{\smallskip}
\end{tabular}
%\label{tabexp1}
\end{center}
\end{table*}

\begin{table*}
\caption{Results of the simulation for a magnetic field strength of $B=10^8$ G and accretion rate $\dot{M}=6\times10^{-13} M_\odot\;\rm yr^{-1}$, but a hollow mound profile. All other parameters are set as in table \ref{tabexp1}.}
\label{tabexp2}%\label{hollowdensity10^8}
\begin{center}
\begin{tabular}{c|c|c|c|c|c|c|c}
\hline \hline \noalign{\smallskip}
Mound Height & Max base density & Accreted mass & $\rho_2/\rho_0$ (Max) & $\rho_2/\rho_0$ (Min) & $\rho_0$ (Max) & $\rho_0$ (Min) & Ellipticity\\
 (m) & ($\rm g\;cm^{-3}$) & ($M_\odot$) &  &  & ($\rm g\;cm^{-3}$) & ($\rm g\;cm^{-3}$)&\\
\hline

0.040 & $1.294\times 10^{7}$ & $1.16\times 10^{-15}$ & 2.1252 & 2.1249 & $4.541\times 10^{5}$ & $2.122\times 10^{2}$ & $3.82\times 10^{-17}$\\
0.046 & $1.596\times 10^{7}$ & $1.63\times 10^{-15}$ & 2.1252 & 2.1249 & $5.601\times 10^{5}$ & $2.619\times 10^{2}$ & $5.38\times 10^{-17}$\\
0.052 & $1.918\times 10^{7}$ & $2.20\times 10^{-15}$ & 2.1252 & 2.1249 & $6.732\times 10^{5}$ & $3.169\times 10^{2}$ & $7.26\times 10^{-17}$\\
0.058 & $2.259\times 10^{7}$ & $2.88\times 10^{-15}$ & 2.1251 & 2.1248 & $7.930\times 10^{5}$ & $3.770\times 10^{2}$ & $9.52\times 10^{-17}$\\
0.064 & $2.618\times 10^{7}$ & $3.68\times 10^{-15}$ & 2.1251 & 2.1247 & $9.191\times 10^{5}$ & $4.431\times 10^{2}$ & $1.21\times 10^{-16}$\\
0.070 & $2.225\times 10^{7}$ & $4.61\times 10^{-15}$ & 2.1251 & 2.1245 & $1.051\times 10^{6}$ & $5.165\times 10^{2}$ & $1.52\times 10^{-16}$\\
0.076 & $3.388\times 10^{7}$ & $5.67\times 10^{-15}$ & 2.1251 & 2.1242 & $1.189\times 10^{6}$ & $5.991\times 10^{2}$ & $1.88\times 10^{-16}$\\
0.082 & $3.797\times 10^{7}$ & $6.89\times 10^{-15}$ & 2.1250 & 2.1239 & $1.333\times 10^{6}$ & $6.933\times 10^{2}$ & $2.28\times 10^{-16}$\\
0.088 & $4.221\times 10^{7}$ & $8.28\times 10^{-15}$ & 2.1249 & 2.1234 & $1.482\times 10^{6}$ & $8.024\times 10^{2}$ & $2.75\times 10^{-16}$\\
0.094 & $4.661\times 10^{7}$ & $9.87\times 10^{-15}$ & 2.1252 & 2.1228 & $1.636\times 10^{6}$ & 1.012 & $3.29\times 10^{-16}$\\
0.100 & $5.114\times 10^{7}$ & $1.16\times 10^{-14}$ & 2.1251 & 2.1220 & $1.795\times 10^{6}$ & 9.016 & $3.90\times 10^{-16}$\\
0.106 & $5.581\times 10^{7}$ & $1.37\times 10^{-14}$ & 2.1251 & 2.1210 & $1.959\times 10^{6}$ & $2.589\times 10$ & $4.61\times 10^{-16}$\\
0.112 & $6.062\times 10^{7}$ & $1.61\times 10^{-14}$ & 2.1250 & 2.1196 & $2.128\times 10^{6}$ & $5.332\times 10$ & $5.44\times 10^{-16}$\\
0.118 & $6.555\times 10^{7}$ & $1.88\times 10^{-14}$ & 2.1249 & 2.1178 & $2.301\times 10^{6}$ & $9.388\times 10$ & $6.42\times 10^{-16}$\\
0.124 & $7.062\times 10^{7}$ & $2.20\times 10^{-14}$ & 2.1249 & 2.1155 & $2.479\times 10^{6}$ & $1.512\times 10^{2}$ & $7.58\times 10^{-16}$\\
0.130 & $7.580\times 10^{7}$ & $2.57\times 10^{-14}$ & 2.1249 & 2.1124 & $2.661\times 10^{6}$ & $2.303\times 10^{2}$ & $8.97\times 10^{-16}$\\
0.136 & unstable &  &  &  &  &  & \\

\hline \hline \noalign{\smallskip}
\end{tabular}
\end{center}
%\label{tabexp2}
\end{table*}

\begin{table*}
\caption{Results of the simulation for a magnetic field strength of $B=10^8$ G and accretion rate $\dot{M}=3 \times 10^{-8} M_\odot\;\rm yr^{-1}$, (approximately the Eddington limit), for a parabolic mound profile. The polar spread is now 60.66 degrees at this high accretion rate.}
\label{tabexp3}%\label{hollowdensity10^8}
\begin{center}
\begin{tabular}{c|c|c|c|c|c|c|c}
\hline \hline \noalign{\smallskip}
Mound Height & Max base density & Accreted mass & $\rho_2/\rho_0$ (Max) & $\rho_2/\rho_0$ (Min) & $\rho_0$ (Max) & $\rho_0$ (Min) & Ellipticity\\
 (m) & ($\rm g\;cm^{-3}$) & ($M_\odot$) &  &  & ($\rm g\;cm^{-3}$) & ($\rm g\;cm^{-3}$)&\\
\hline

0.24 &$1.901\times 10^8$ &$ 1.40\times 10^{-12}$ & 2.0987 & 1.2938 & $ 8.9918\times 10^7$ &$3.8860\times 10^4$& $3.52\times 10^{-13}$\\
0.25 &$2.022\times 10^8$ &$ 1.58\times 10^{-12}$ & 2.0962 & 1.2938 & $ 9.5595\times 10^7$ &$4.1953\times 10^4$ &$3.99\times 10^{-13}$\\
0.26 &$2.144\times 10^8$ &$ 1.77\times 10^{-12}$ & 2.0931 & 1.2938 & $ 1.0139\times 10^8$ &$4.5307\times 10^4$ &$4.53\times 10^{-13}$\\
0.27 &$2.269\times 10^8$ &$ 1.98\times 10^{-12}$ & 2.0894 & 1.2938 & $ 1.0729\times 10^8$ &$4.8980\times 10^4$ &$5.16\times 10^{-13}$\\
0.28 &$2.396\times 10^8$ &$ 2.23\times 10^{-12}$ & 2.0849 & 1.2938 & $ 1.1331\times 10^8$ &$5.3043\times 10^4$ &$5.91\times 10^{-13}$\\
0.29 &$2.526\times 10^8$ &$ 2.52\times 10^{-12}$ & 2.0793 & 1.2907 & $ 1.1943\times 10^8$ &$5.7600\times 10^4$ &$6.83\times 10^{-13}$\\
0.30 &$2.657\times 10^8$ &$ 2.85\times 10^{-12}$ & 2.0724 & 1.2733 & $ 1.2566\times 10^8$ &$6.2793\times 10^4$ &$7.99\times 10^{-13}$\\
0.31 &$2.791\times 10^8$ &$ 3.26\times 10^{-12}$ & 2.0635 & 1.2331 & $ 1.3199\times 10^8$ &$6.8832\times 10^4$ &$9.51\times 10^{-13}$\\
0.32 & unstable & &  &  & & &\\
\hline \hline \noalign{\smallskip}
\end{tabular}
\end{center}
%\label{tabexp3}
\end{table*}

\begin{table*}
\caption{Results of the simulation for a magnetic field strength of $B=10^{10}$ G and accretion rate $\dot{M}= 3\times 10^{-8} M_\odot\;\rm yr^{-1}$, (approximately the Eddington limit), for a parabolic mound profile. The polar spread is 16 degrees.}
\label{tabhighMB}
\begin{center}
\begin{tabular}{c|c|c|c|c|c|c|c}
\hline \hline \noalign{\smallskip}
Mound Height & Max base density & Accreted mass & $\rho_2/\rho_0$ (Max) & $\rho_2/\rho_0$ (Min) & $\rho_0$ (Max) & $\rho_0$ (Min) & Ellipticity\\
 (m) & ($\rm g\;cm^{-3}$) & ($M_\odot$) &  &  & ($\rm g\;cm^{-3}$) & ($\rm g\;cm^{-3}$)&\\
\hline

1.05 &$1.740\times 10^9$ &$ 4.59\times 10^{-12}$ & 2.2315 & 2.1492 & $ 7.0045\times 10^7$ &$6.0412\times 10^2$ &$1.05\times 10^{-13}$\\
1.10 &$1.866\times 10^9$ &$ 5.21\times 10^{-12}$ & 2.2308 & 2.1492 & $ 7.5106\times 10^7$ &$1.0863\times 10^3$ &$1.21\times 10^{-13}$\\
1.15 &$1.994\times 10^9$ &$ 5.90\times 10^{-12}$ & 2.2302 & 2.1492 & $ 8.0285\times 10^7$ &$1.7004\times 10^3$ &$1.38\times 10^{-13}$\\
1.20 &$2.126\times 10^9$ &$ 6.67\times 10^{-12}$ & 2.2297 & 2.1492 & $ 8.5576\times 10^7$ &$2.4445\times 10^3$ &$1.58\times 10^{-13}$\\
1.25 &$2.260\times 10^9$ &$ 7.52\times 10^{-12}$ & 2.2292 & 2.1492 & $ 9.0983\times 10^7$ &$3.3190\times 10^3$ &$1.80\times 10^{-13}$\\
1.30 &$2.397\times 10^9$ &$ 8.49\times 10^{-12}$ & 2.2287 & 2.1492 & $ 9.6495\times 10^7$ &$4.3264\times 10^3$ &$2.07\times 10^{-13}$\\
1.35 &$2.537\times 10^9$ &$ 9.59\times 10^{-12}$ & 2.2283 & 2.1487 & $ 1.0211\times 10^8$ &$5.4724\times 10^3$ &$2.39\times 10^{-13}$\\
1.40 &$2.679\times 10^9$ &$ 1.09\times 10^{-11}$ & 2.2341 & 2.1473 & $ 1.0784\times 10^8$ &$3.6423\times 10^1$ &$2.78\times 10^{-13}$\\
1.45 &$2.824\times 10^9$ &$ 1.23\times 10^{-11}$ & 2.2330 & 2.1445 & $ 1.1367\times 10^8$ &$2.0099\times 10^2$ &$3.27\times 10^{-13}$\\
1.50 &$2.971\times 10^9$ &$ 1.41\times 10^{-11}$ & 2.2321 & 2.1399 & $ 1.1960\times 10^8$ &$4.9756\times 10^2$ &$3.91\times 10^{-13}$\\
1.55 &$3.121\times 10^9$ &$ 1.62\times 10^{-11}$ & 2.2314 & 2.1340 & $ 1.2563\times 10^8$ &$9.2825\times 10^2$ &$4.70\times 10^{-13}$\\
1.60 & unstable & & & & & &\\
\hline \hline \noalign{\smallskip}
\end{tabular}
\end{center}
%\label{tabhighMB}
\end{table*}

\begin{figure}
\includegraphics[width=1\columnwidth]{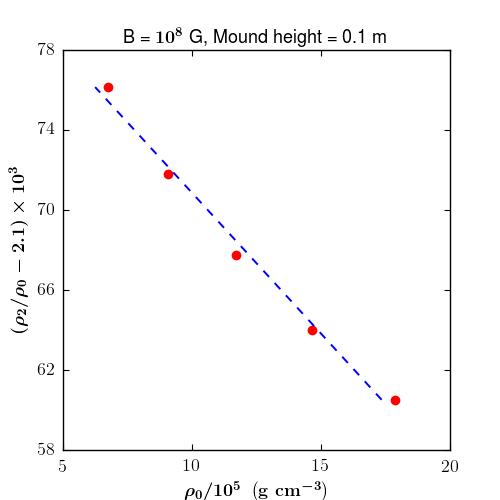}
\caption{Linear fitting for mound of height 0.1m formed for magnetic field strength of $B=10^8$ G and a parabolic mound profile as shown in figure \ref{ratioP}. The linear fitting is done for the points at higher $\rho_0 $ which show a linear behavior with $\rho_2/\rho_0 $. Note that the ticks have been chosen in order to have a convenient labeling.}\label{linearfitpara01}
\end{figure}

We see that, in general, the densities at the base of the crust are lower than that of many of the reaction layers we consider, and the ratio $ \rho_2/ \rho_0$ appears to be linearly decreasing with density. This suggests that we may extrapolate our results to higher densities by means of a fit of the dependence of $\rho_2/\rho$ on $\rho$. We choose a linear fit, mostly for simplicity, as this appears to be a good approximation to our data. In fact we do not have enough data points close to the base of the mountain and spread over a large enough dynamical range to meaningfully perform a logarithmic fit, or fit a power-law with varying index. We can further assess this approximation with a simple analysis of the forces acting on the system. Let us consider a plane parallel section of a neutron star, where gravity $g$ is directed in the negative $z$ direction. The Alfven wavelength, i.e. the length-scale at which magnetic stresses become the dominant force in the system \citet{HS09}
\be
\lambda_a=H \sqrt{gH}\left(\frac{B_z}{\sqrt{2\pi\rho}}\right)^{-1}\label{spitkovsky}
\ee
with $B_z$ the magnetic field along the $z$ axis and the scale-height $H=p/\rho g$. We can now estimate the quadrupolar mass distribution by assuming that the accreted material can only spread within $\lambda_a$ and make the approximation that $\lambda_a$ can be taken to be also the curvilinear length-scale over which matter is confined at a given radius \citep{2011ApJ...740L...8C}. Then the $l=2$ component of the density is given by
\be
\rho_2\approx 2\pi \int_0^{(\lambda_a/R)} \rho Y_{20} \sin\theta d\theta\propto\sin^2(\lambda_a/R) \cos(\lambda_a/R)
\ee
which as expected vanishes for $\lambda_a/R=0$ and $\lambda_a/R=\pi/2$, i.e. when matter has spread to the whole surface.
If we expand close to $\lambda_a/R=\pi/2$ we see that for high densities $\rho_2/\rho\propto (\pi/2- \lambda_a/R)$. Assuming a polytropic equation of state, we have from (\ref{spitkovsky})
\be
\frac{\rho_2}{\rho}\propto \rho^{3\Gamma/2-1}
\ee
which gives $\frac{\rho_2}{\rho}\propto \rho^{3/2}$ for our $\Gamma=5/3$ equation of state, thus indicating a powerlaw behaviour that is not, however, too far from our linear approximation. Note that the above analysis is valid for weak fields in which matter can spread significantly. An approximate estimate of the confinement radius can be found by equating the magnetic Lorentz force to the pressure gradient, as derived in %If the magnetic pressure is strong compared to the gas pressure the accreted material can effectively be confined by the magnetic field inside a radius.
\citet{BB98}
\be
R_c\leq \frac{8\pi Hp}{B^2}
\ee
Repeating the same analysis as above, we find that for strong fields we may expect $\rho_2/\rho\propto\rho^{(2\Gamma-1)}$, which for our polytrope gives $\rho_2/\rho\propto\rho^{7/3}$. In practice in the presence of significant bending of the field lines due to large accreted masses, the ratio $\rho_2/\rho_0$ obtained from our simulations deviates significantly from a linear relation, as can be seen in figures \ref{ratioP} and \ref{ratioH}, and the effects of the lower boundary play a significant role. In these cases, we do not attempt to extrapolate our results to higher densities, but rather extrapolate only when the linear fit is justified by the data. An example of a linear fit for parabolic mound profile is shown in figure \ref{linearfitpara01}

\begin{figure}
\includegraphics[width=1\columnwidth]{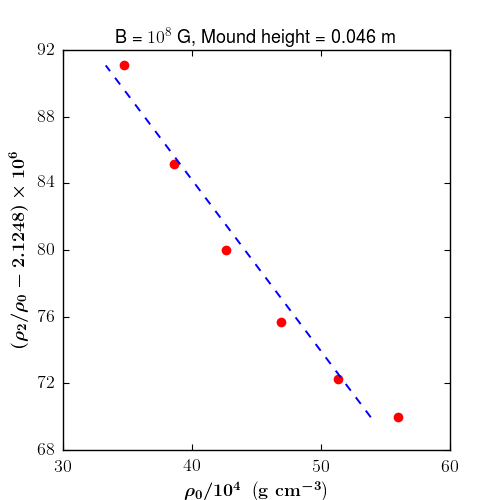}
\caption{Linear fitting for mound of height 0.046m formed for magnetic field strength of $B=10^8$ G and a hollow mound profile.}\label{linearfithollow46}
\end{figure} 

\begin{figure*}
\includegraphics[width=1\columnwidth]{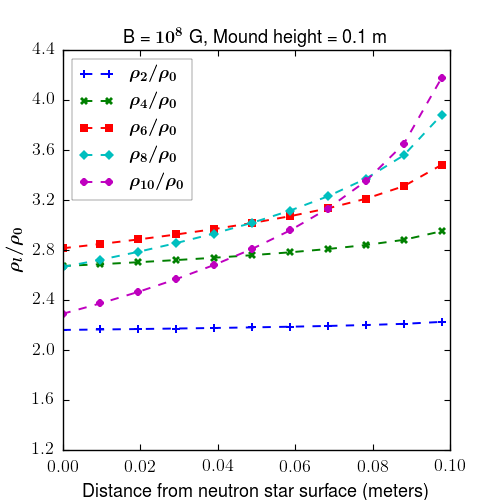}
\includegraphics[width=1\columnwidth]{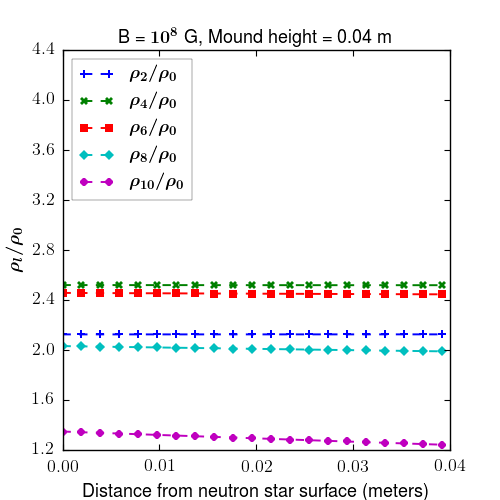}
\caption{Left panel: Ratios of higher harmonics of the density distributions $\rho_l/\rho_0$ versus distance from the surface for a parabolic (left) and hollow (right) mound profile. In the parabolic case it can be seen that higher harmonics of the density distribution fall off faster than the $l=2$ harmonic with increasing depth, justifying our approximation that this is the main contribution at high densities. For the hollow case this approximation is questionable, as the density distribution still requires the contribution of higher order harmonics to be described also at high densities.}\label{ratios}
\end{figure*}

\begin{figure*}
\includegraphics[width=\columnwidth]{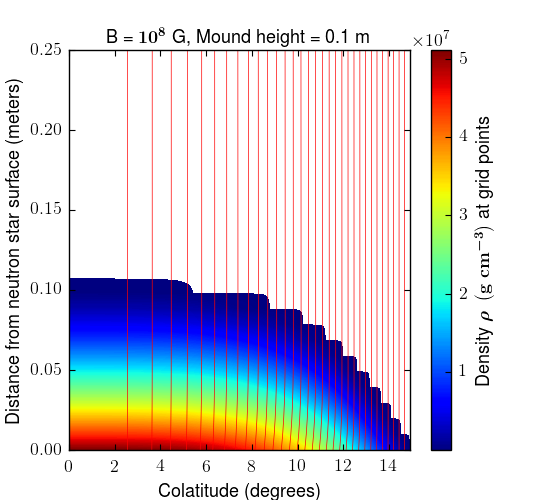}\includegraphics[width=\columnwidth]{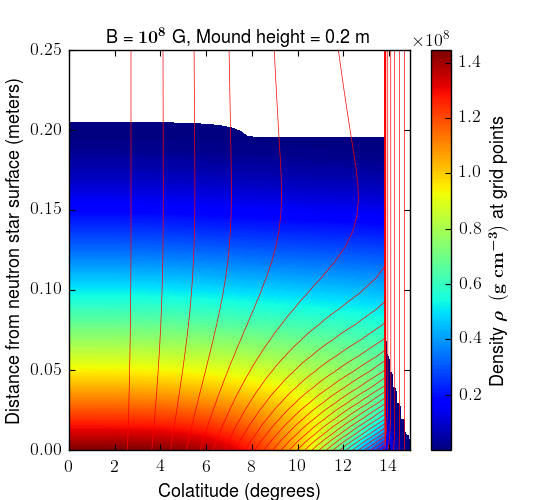}\\
\includegraphics[width=\columnwidth]{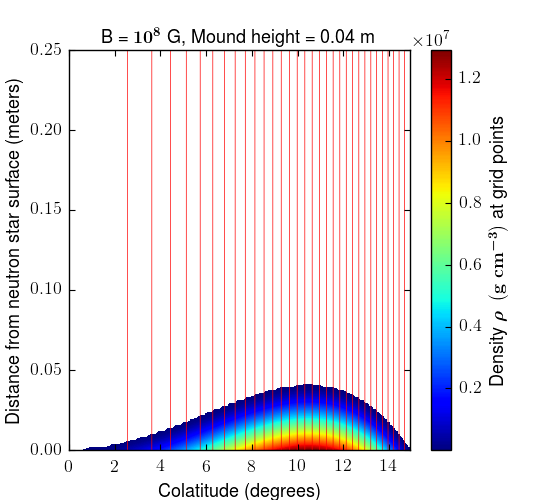}\includegraphics[width=\columnwidth]{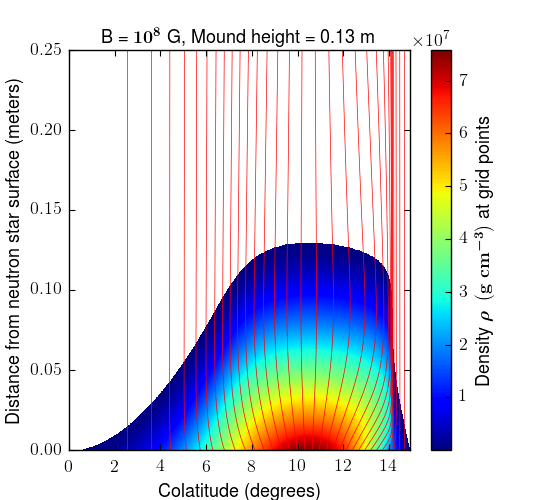}
\caption{Density distributions for a parabolic mound in the top panel, for two mound heights, 0.1 m and 0.2 m and for a hollow mound in the bottom panel, for mound heights of 0.04 m and 0.13 m. Other parameters are as in tables \ref{tabexp1} and \ref{tabexp2}, although note that for numerical reasons a maximum grid height of 5 m was used for parabolic mounds, and of 1 m for hollow mounds. The magnetic axis being the symmetry axis of our basis of spherical harmonics, the parabolic mound remains symmetric around the pole and only begins to be strongly deformed for higher mounds. The hollow mound is much more asymmetric as expected, and requires a larger number of spherical harmonics to be reconstructed for all mound heights.}
\label{compare}
\end{figure*}

Overall, our main justification for a linear fit is, thus, numerical. Although a power-law fit of the form $\rho_2/\rho=\rho_0^k$ may be more justified, the analysis above suggests that $1\lesssim k \lesssim 2$, and is generally valid only for small mounds in which the field lines are not bent significantly and matter does spread far from the polar caps. Given the large physical uncertainties in the model for larger mounds, and small dynamical range of the data we can fit over, we choose a linear fit as suggested by the data. If, indeed, the deformation $\rho_2/\rho$ scales with a somewhat higher power of density, closer to $k \approx 2$, then our results are to be interpreted as upper limits.

We also test the validity of our extrapolation procedure and of the perturbative expansion by plotting also the ratio between higher $l$ multipoles of density and the spherical component, i.e. the ratios $\rho_l/\rho_0$, up to $l=10$. The results are given in figure \ref{ratios} for both the parabolic and hollow mound profiles. We see that for a parabolic mound, in general higher multipoles decrease faster as density increases, validating our approximation that at higher densities at which the reactions in table \ref{deep} occur, the density distribution is mainly spherical with small deviations from axisymmetry. For a hollow mound profile the situation is different, as the density distribution is still highly asymmetric at the base of our grid. This is to be expected given the 'hollow' nature of the profile that naturally leads to a roughly bimodal density distribution. This can be seen clearly by comparing the density distributions in figure \ref{compare}, where it is clear that even for small accreted masses and mound heights of 0.1 m for a parabolic mound, the density distribution has a single peak centered around the axis, while for a hollow mound the peak is off center, and the density distribution is highly non-spherical.
As we will see below this is not only a more realistic setup, but also the most interesting for gravitational wave emission as it naturally leads to higher asymmetries and `mountains'. It is also, however, the setup for which our extrapolation is less reliable and our results are thus likely to only be lower limits, as they neglect non-linear effects and the contribution from higher values of $l$. A consistent, non-linear analysis of the problem is clearly needed in the future to assess this problem quantitatively.

The results for the extrapolation are given in tables \ref{ext1}, \ref{ext2} and \ref{fitEDD} where we show the results of our fitting procedure, for which we assume a linear dependence between $\rho_0$ and the ratio $\rho_2/\rho_0$  of the form:
\begin{equation}
\rho_0=A\frac{\rho_2}{\rho_0}+B \label{linearex}
\ee
In general we see that for small amounts of accreted mass, comparable to what a weakly magnetised and slowly accreting source such as PSR J1023+0038 would accrete in a few days of outburst, the quadrupolar deformation, for filled accretion mounds, vanishes at densities of a few times $\rho\approx 10^8\rm g\;cm^{-3}$. Deformations are present at higher densities for higher magnetic fields, which limit the spread of matter on the star, but generally are not present above $\rho\approx 10^9 \rm g\;cm^{-3}$. 
The same is true also for sources accreting close to the Eddington limit, which suggests that even in persistently accreting sources, which are prime targets for gravitational wave detection \citep{Haskell15}, one cannot obtain large thermal mountains.

For the more realistic case of a hollow mound, however, deformations can persist up to $\rho\approx 10^{10}\rm g\;cm^{-3}$ in weakly magnetised neutron stars. For strong magnetic fields, however, the the ratio $\rho_2/\rho_0$ is a strongly non-linear function of $\rho$, as is also the case for large accreted masses, for which the mound height approaches it's critical value. 
In all cases where the extrapolation to high densities is possible, i.e. for weak fields and low accretion rates (as is the case for PSR J1023+0038) deformations of the order of $\delta T_q/\Delta T\approx 0.01$ are not possible at densities $\rho \approx 10^{12}\rm g\;cm^{-3}$, such as those of the most energetic pycno-nuclear reactions considered by \citet{Haskell17}.

\begin{table*}
\caption{Results of the linear extrapolation of the ratio $\rho_2/\rho_0$ at higher densities, following the relation in (\ref{linearex}), $\rho_0=A\frac{\rho_2}{\rho_0}+B$, for the parabolic mound models in table \ref{tabexp1}. It can be concluded from the $\rho_0$ value for $\rho_2/\rho_0 = 1\%$ in last column that the ratio $\rho_2/\rho_0$ will be already less than $1\%$ for densities well below $10^9\rm g\;cm^{-3}$ and thus at densities lower than those of the first deep crustal heating reactions in table \ref{deep}.}
\label{ext1}
\begin{center}
\begin{tabular}{c|c|c|c|c|c}
\hline \hline \noalign{\smallskip}
Accreted mass & Mound Height & A(slope) & B(intercept) & $\rho_0$ for ($\rho_2/\rho_0 = 1\%$)\\
 ($M_\odot$)&(m) &($\rm g\;cm^{-3}$) &($\rm g\;cm^{-3}$) & ($\rm g\;cm^{-3}$) \\
\hline
$2.44\times 10^{-15}$& 0.05 & $-2.382\times 10^{7}$ & $5.209\times 10^{7}$ &  $5.2053\times 10^{7}$\\
$1.23\times 10^{-14}$&0.10 & $-7.094\times 10^{7}$ & $1.550\times 10^{8}$ &  $1.5493\times 10^{8}$\\
$3.48\times 10^{-14}$&0.15 & $-1.857\times 10^{8}$ & $4.043\times 10^{8}$ &  $4.0402\times 10^{8}$\\
$9.34\times 10^{-14}$&0.20 & Not linear & & &\\
\hline \hline \noalign{\smallskip}
\end{tabular}
\end{center}
\end{table*}

\begin{table*}
\caption{Results of the linear extrapolation as in table (\ref{ext1}), but for the hollow mound models in table \ref{tabexp2}. The ratio $\rho_2/\rho_0$ remains sizeable at higher densities than for a parabolic mount, up to densities at which reactions in table \ref{deep} can occur.}
\label{ext2}
\begin{center}
\begin{tabular}{c|c|c|c|c|c}
\hline \hline \noalign{\smallskip}
Accreted mass & Mound Height & A(slope) & B(intercept) & $\rho_0$ for ($\rho_2/\rho_0 = 1\%$)\\
 ($M_\odot$)&(m) &($\rm g\;cm^{-3}$) & ($\rm g\;cm^{-3}$)& ($\rm g\;cm^{-3}$) \\
\hline
$1.16\times 10^{-15}$&0.040 & $-4.111\times 10^{9}$ & $8.735\times 10^{9}$ &   $8.7321\times 10^{9}$\\
$1.31\times 10^{-15}$&0.042 & $-5.231\times 10^{9}$ & $1.112\times 10^{10}$ &  $1.1113\times 10^{10}$\\
$1.46\times 10^{-15}$&0.044 & $-7.136\times 10^{9}$ & $1.516\times 10^{10}$ & $1.5155\times 10^{10}$\\
$1.63\times 10^{-15}$&0.046 & $-9.721\times 10^{9}$ & $2.065\times 10^{10}$ &  $2.0644\times 10^{10}$\\
$1.81\times 10^{-15}$&0.048 & Not linear & & &\\
\hline \hline \noalign{\smallskip}
\end{tabular}
\end{center}
\end{table*}

\begin{table*}
\caption{Results of the linear extrapolation as in table (\ref{ext1}), for the parabolic mound models in table \ref{tabexp3} with accretion rates close to the Eddington limit and $B=10^{8}$ G. As can be seen the ratio $\rho_2/\rho_0$ is larger at higher densities than for the weaker accretion rates considered in table (\ref{ext1}), but vanishes just before reaching the densities at which reactions in table \ref{deep} can occur.}
\label{fitEDD}
\begin{center}
\begin{tabular}{c|c|c|c|c|c}
\hline \hline \noalign{\smallskip}
Accreted mass & Mound Height & A(slope) & B(intercept) & $\rho_0$ for ($\rho_2/\rho_0 = 1\%$)\\
 ($M_\odot$)&(m) & ($\rm g\;cm^{-3}$)&($\rm g\;cm^{-3}$) & ($\rm g\;cm^{-3}$) \\
\hline
$1.40\times 10^{-12}$&0.24 & $-3.735\times 10^{8}$ & $5.715\times 10^{8}$ &   $5.7118\times 10^{8}$\\
$1.58\times 10^{-12}$&0.25 & $-4.381\times 10^{8}$ & $6.610\times 10^{8}$ &  $6.6047\times 10^{8}$\\
$1.77\times 10^{-12}$&0.26 & $-5.111\times 10^{8}$ & $3.800\times 10^{8}$ & $7.5983\times 10^{8}$\\
$1.98\times 10^{-12}$&0.27 & $-6.590\times 10^{8}$ & $5.176\times 10^{8}$ &  $9.5642\times 10^{8}$\\
$2.23\times 10^{-12}$&0.28 & Not linear & & &\\
\hline \hline \noalign{\smallskip}
\end{tabular}
\end{center}
\end{table*}

\begin{table*}
\caption{Results of the linear extrapolation of the ratio $\rho_2/\rho_0$ at higher densities, following the relation in eq.(\ref{linearex}), $\rho_0=A\frac{\rho_2}{\rho_0}+B$, for the parabolic mound models in table \ref{tabhighMB} with accretion rate at the Eddington limit and $B=10^{10}$ G.}
\label{linefit10e10}
\begin{center}
\begin{tabular}{c|c|c|c|c|c}
\hline \hline \noalign{\smallskip}
Accreted mass & Mound Height & A(slope) & B(intercept) & $\rho_0$ for ($\rho_2/\rho_0 = 1\%$)\\
($M_\odot$)&(m) & & & ($\rm g\;cm^{-3}$) \\
\hline
$ 4.59\times 10^{-12}$ & 1.05 & $-3.286\times 10^{9}$ & $7.129\times 10^{9}$ &  $7.1258\times 10^{9}$\\
$ 5.21\times 10^{-12}$ & 1.10 & $-3.785\times 10^{9}$ & $8.207\times 10^{9}$ &  $8.2034\times 10^{9}$\\
$ 5.90\times 10^{-12}$& 1.15 & $-4.437\times 10^{9}$ & $9.665\times 10^{9}$ &  $9.6605\times 10^{9}$\\
$ 6.67\times 10^{-12}$ & 1.20 & $-5.464\times 10^{9}$ & $1.182\times 10^{10}$ &  $1.1821\times 10^{10}$\\
$ 7.52\times 10^{-12}$ & 1.25 & Not linear & & &\\
\hline \hline \noalign{\smallskip}
\end{tabular}
\end{center}
\end{table*}

\section{Induced quadrupole}

Following the results of the previous section, it is clear that in most cases no significant deformation persists up to densities at which reactions can occur. For parabolic mounds this is only the case for high accretion rates and strong magnetic fields. For weaker fields, of the order of $B\approx 10^8$ G, as commonly inferred in LMXBs, only the hollow mounds allow significant deformation at the densities at which the reactions in table \ref{deep} take place. 

We consider this case first, and calculate the quadrupole for such a hollow mound. Despite not releasing as much energy as reactions in deeper layers, it is clear that shallow layers play a role, as deformations can be large at those depths. As an example let us consider the case of PSR J1023+0038, for which the increase in spin-down rate during outburst can be explained by a quadrupole of \citet{Haskell17}
\be
Q_{22}=4.4\times 10^{35} \;\;\mbox{g cm$^2$}
\ee
where we assume a moment of inertia $I=10^{45}$ g cm$^2$. We can compare this to the quadrupole of our model with an elliptical mound, $B=10^8$ G and an accreted mass of $4\times 10^{18}$ g (approximately the amount of mass PSR J1023+0038 would accrete in the first day or two of outburst).

The maximum density at the base of our grid is $\rho_0=5.601\times10^5 \rm g\;cm^{-3}$ and the corresponding $\rho_2/\rho_0=2.1249$. The relation between $\rho_2/\rho_0$ and $\rho$ appears linear, as can be seen in figure \ref{linearfithollow46} so we extrapolate our results with the linear fit $\rho_0=-9.721\times 10^9 (\rho_2/\rho_0)+2.065\times 10^{10}\rm g\;cm^{-3}$. From this, and using the data from table \ref{deep} in equation (\ref{total}), we obtain, assuming a background temperature $T=10^7$ K, $A=85$ and $Z/A=0.4$.
\be
Q_T\approx-10^{31}  \;\mbox{g cm$^2$}
\ee

In comparison the  mass quadrupole induced by the matter confined by the magnetic field on the surface is $Q\approx 10^{29}$\;\mbox{g cm$^2$} and negligible. We see that the quadrupole is well below what is needed to explain the observations. One may object that a larger amount of mass is accreted during an outburst, $\Delta M\approx 10^{20}$ g, and despite our models showing that the mound is unstable for such values, numerical time-evolutions suggest that the mound height saturates even if additional mass is added \citep{Vigelius}. We may thus extrapolate to larger values of the accreted mass. $Q_T$ grows linearly with accreted mass, so even for accreted masses of order $10^{21}$ g, which is generally more than accreted during an outburst the quadrupole would not be large enough to explain the spin-down of PSR J1023+0038.

Let us now turn our attention to the high accretion rate, where the star is accreting close to the Eddington limit and the background field is $B=10^{10}$ G. In this case density perturbations extend to the first reaction layer in table \ref{deep}. If we assume in this case, a hotter star with an internal temperature of $T=10^8$ K, from the results in table \ref{linefit10e10} we obtain a quadrupole of 
\be
Q_T\approx-4\times 10^{34}  \;\mbox{g cm$^2$}
\ee
which corresponds to an ellipticity of $\epsilon\approx 5\times 10^{-11}$ and a signal that is likely to be too weak for detection with current ground based interferometers \citep{Ligo2, Ligo1, Ligo3}.

We note, however, that several studies of cooling X-ray transients have found that the cooling curves cannot be reproduced unless a shallow heating source with $Q_M\approx 1-10$ MeV is included at densities $\rho\approx 10^{8}-10^{10}\rm g\;cm^{-3}$ \citep{Shallow1, Shallow2, Shallow3, Shallow4}. If an additional shallow heating layer is added to the reactions in table \ref{deep}, at $\rho=10^9\rm g\;cm^{-3}$ and with $Q_M\approx 5$ MeV, using the BSk20 EoS \citep{Goriely}, and take an accreted mass of $\Delta M\approx 10^{20}$ g, 
we obtain $Q_T\approx - 8\times 10^{34}$ g cm$^2$. A strong shallow heating source is thus consistent with a scenario in which gravitational waves spin down PSR J1023+0038 during outburst. Furthermore if such hypothetical shallow heating sources are present in systems accreting close to the Eddington limit, with a magnetic field of $B=10^{10}$ G, then one would have $Q_T\approx -3\times 10^{38}$ g cm$^2$, corresponding to an ellipticity of $\epsilon\approx 4\times 10^{-7}$ and potentially detectable in the near future by Advanced LIGO and Virgo from known persistently accreting neutron stars in LMXBs.

The above is an estimate of the quadrupole that can be formed during a single outburst. If the quadrupole is indeed sourced uniquely by temperature asymmetries, we expect it to be washed out on a thermal timescale for the crust, which for a capture layer at a pressure $P_{30}$ in units of $10^{30}$ erg/cm$^3$, is given by \citet{BBR98}
\be
\tau_{th}=0.2 P_{30}^{3/4} \mbox{yrs}
\ee
so that a shallow `thermal' mountain as we have considered would be washed out entirely a few months after the outburst.
However, as a consequence of the outburst, compositional asymmetries will build up as the reactions proceed, and may be accumulated over several outbursts \citep{UCB}, allowing to build up a mountain over several outbursts. In fact, the recent analysis of the spin-distribution of accreting neutron star by \citet{Gittins} shows that population synthesis models including a quadrupole from a mountain built up over many outbursts fit the data better than the models which assume that a mountain is built up over a single outburst and then washed away.

\section{Conclusions}

We have calculated numerical models of the outer layers of an accreting magnetised neutron star, using the numerical Grad-Shafranov code of \citet{code12}. The density profile in these models has been expanded in spherical harmonics and the ratio between the quadrupolar and spherical density perturbations $\delta \rho_2/\delta\rho_0$, extrapolated linearly to higher densities. We find that for a filled mound with a parabolic profile function for the height, the quadrupolar density perturbation vanishes at densities lower than $\rho\approx 10^9\rm g\;cm^{-3}$ for field strengths of $B\approx 10^8$ G, typical of  neutron stars in LMXBs, for all accretion rates, and only for higher magnetic fields of $B=10^{10}$ G do we have significant perturbations up to densities of $\rho\approx 10^{10}\rm g\;cm^{-3}$ . For a more realistic hollow accretion profile \citep{inst1} the quadrupolar density perturbation persists up to densities at which deep crustal heating reactions occur \citep{Fantina18}.
For these models we calculate the quadrupolar temperature perturbations due to the reactions, following the formulation of \citet{UCB}, to estimate the size of the additional mass quadrupole that would be induced.
We consider in particular the case of PSR J1023+0038, for which it has been suggested that a gravitational wave mountain may be created during an accretion outburst \citep{Haskell17}, explaining the enhanced spin-down of the star. For standard reactions  we obtain quadrupoles of the order
\be
Q_T\approx 10^{31}-10^{33}\;\mbox{g cm$^2$}
\ee
for this source, which is not sufficient to explain the additional spin-down. 

However the cooling curves of many accreting sources reveal the need to add additional shallow heating sources at densities $\rho\lesssim 10^{10}\rm g\;cm^{-3}$ \citep{Shallow1, Shallow2, Shallow3, Shallow4}. Indeed if we add a strong shallow heating source which releases $Q=5$ MeV of heat at a density of $\rho=10^9\rm g\;cm^{-3}$, we find that one can build, over the outburst, a mountain corresponding to a quadrupole
\be
Q_T\approx -8 \times 10^{34}\;\mbox{g cm$^2$}
\ee
which would explain the increased spin-down of PSR J1023+0038 during outburst. Furthermore, for a system accreting persistently at the Eddington rate and with a magnetic field of $B=10^{10}$ G, the quadrupole could be as large as
\be
Q_T\approx -3 \times 10^{38}\;\mbox{g cm$^2$}
\ee
and lead to a GW signal that is potentially detectable by Advanced LIGO and Virgo for known persistently accreting LMXBs, although even at this level the search would be challenging \citep{Watts08}.

We stress, however, that our estimates rely on extrapolations at densities higher than those of our computational domain, and that the large quadrupoles we obtain, larger than those induced by the magnetic deformations themselves, indicate that treating this effect as a linear perturbation is likely to be an inadequate approach. By limiting ourselves to cases in which the relation between the quadrupolar and spherical density perturbations is linear, we are limiting ourselves to cases in which there is no substantial line bending, and the magnetic deformations are also small. For larger deformations, MHD instabilities may be present \citep{2001L,inst1, inst2} and may limit the growth of mounds beyond the heights considered in this work. In reality the true values of the mass quadrupole may thus be close to the lower limits we compute. Furthermore, the equation of state we use does not consistently model all the regions of the crust we consider, and in particular is inadequate for low densities in the outer layers, where one should account for degenerate electrons. In the current work we considered only a single polytropic EOS, adequate to describe degenerate neutrons, as this allowed us to obtain models for higher accreted masses. Polytropic models for degenerate electron EOSs (both relativistic and non-relativisitc) were also explored, but displayed a non-linear relation between the quadrupolar and spherical density perturbations already for values of the accreted mass much lower than those of interest, and did not allow to extrapolate the behaviour of the quadrupole to higher densities.  Future work will aim to consistently model magnetic deformations and crustal reactions, with a realistic equation of state, in order to accurately estimate the induced quadrupole. 

Finally we note that we have only considered deformations of the magnetic field and corresponding density profile. However the early evolution of the magnetic field due to Hall drift can lead to small scale structures that can persist for timescales greater than $10^6$ yrs, leading to sizable deformations \citet{Geppert}. If these structures can persist, even partially, in the crust of accreting neutron stars, they will lead to additional deformations at high density and deform deep reaction layers, possibly allowing for larger mountains and stronger gravitational wave emission.

\section*{Acknowledgements}
%\begin{acknowledgments}
We acknowledge support from the Polish National Science Centre grant SONATA BIS 2015/18/E/ST9/00577. Partial support comes from PHAROS, COST Action CA16214.  TB and NS acknowledges support of the FNP grant TEAM/2016-3/19. BH wishes to thank J.~L.~Zdunik for interesting discussions and for sharing a preliminary version of the data in table (\ref{deep}). The authors are very grateful to D.I.~Jones for useful comments and suggestions. NS wishes to thank N. ~Andersson for insightful discussion and comments, and D.~Bhattacharya for guidance and assistance in the initial development of the code.  This document has been assigned LIGO document number  P1900232
%\end{acknowledgments}
\bibliographystyle{mnras}
\bibliography{Mount} %,modrefs,psrrefs,crossrefs}

\end{document}